\documentclass[aps,prd,twocolumn,superscriptaddress,showpacs]{revtex4-1}

\usepackage{amsmath}
\usepackage{graphicx}
\usepackage{pifont}
\usepackage{array}
\usepackage{rotate}
\usepackage{vmargin}
\usepackage{lineno}
\usepackage{fancyhdr}
\usepackage{color}
\usepackage{url}
\usepackage{wasysym}
\usepackage{textcomp}

\newcommand{\MADGRAPH}{\textsc{MadGraph}}

\begin{document}


\sloppy


\newpage

\title{Search for magnetic monopoles with the MoEDAL forward trapping detector in 2.11~fb$^{-1}$ of 13 TeV proton-proton collisions at the LHC}

\author{B.~Acharya}
\affiliation{Theoretical Particle Physics \& Cosmology Group, Physics Dept., King's College London, UK}
\affiliation{International Centre for Theoretical Physics, Trieste, Italy}   

\author{J.~Alexandre}
\affiliation{Theoretical Particle Physics \& Cosmology Group, Physics Dept., King's College London, UK}

\author{S.~Baines}
\affiliation{Theoretical Particle Physics \& Cosmology Group, Physics Dept., King's College London, UK}

\author{P.~Benes}
\affiliation{IEAP, Czech Technical University in Prague, Czech~Republic}

\author{B.~Bergmann}
\affiliation{IEAP, Czech Technical University in Prague, Czech~Republic}

\author{J.~Bernab\'{e}u}
\affiliation{IFIC, Universitat de Val\`{e}ncia - CSIC, Valencia, Spain}

\author{A.~Bevan}
\affiliation{School of Physics and Astronomy, Queen Mary University of London, UK}

\author{H.~Branzas}
\affiliation{Institute of Space Science, Bucharest - M\u{a}gurele, Romania}

\author{M.~Campbell}
\affiliation{Experimental Physics Department, CERN, Geneva, Switzerland}

\author{L.~Caramete}
\affiliation{Institute of Space Science, Bucharest - M\u{a}gurele, Romania}

\author{S.~Cecchini}
\affiliation{INFN, Section of Bologna, Bologna, Italy}


\author{M.~de~Montigny}
\affiliation{Physics Department, University of Alberta, Edmonton, Alberta, Canada}


\author{A.~De~Roeck}
\affiliation{Experimental Physics Department, CERN, Geneva, Switzerland}

\author{J.~R.~Ellis}
\affiliation{Theoretical Particle Physics \& Cosmology Group, Physics Dept., King's College London, UK}
\affiliation{National Institute of Chemical Physics \& Biophysics, R{\" a}vala 10, 10143 Tallinn, Estonia and Theoretical Physics Department, CERN, Geneva, Switzerland}

\author{M.~Fairbairn}
\affiliation{Theoretical Particle Physics \& Cosmology Group, Physics Dept., King's College London, UK}

\author{D.~Felea}
\affiliation{Institute of Space Science, Bucharest - M\u{a}gurele, Romania}


\author{M.~Frank}
\affiliation{Department of Physics, Concordia University, Montr\'{e}al, Qu\'{e}bec,  Canada}

\author{D.~Frekers}
\affiliation{Physics Department, University of Muenster, Muenster, Germany}

\author{C.~Garcia}
\affiliation{IFIC, Universitat de Val\`{e}ncia - CSIC, Valencia, Spain}



\author{J.~Hays}
\affiliation{School of Physics and Astronomy, Queen Mary University of London, UK}

\author{A.~M.~Hirt}
\affiliation{Department of Earth Sciences, Swiss Federal Institute of Technology, Zurich, Switzerland -- Associate member}

\author{J.~Janecek}
\affiliation{IEAP, Czech Technical University in Prague, Czech~Republic}



\author{D.-W.~Kim}
\affiliation{Physics Department, Gangneung-Wonju National University, Gangneung, Republic of Korea}

\author{K.~Kinoshita}
\affiliation{Physics Department, University of Cincinnati, Cincinnati, Ohio, USA}

\author{A.~Korzenev}
\affiliation{Section de Physique, Universit\'{e} de Gen\`{e}ve, Geneva, Switzerland}

\author{D.~H.~Lacarr\`ere}
\affiliation{Experimental Physics Department, CERN, Geneva, Switzerland}

\author{S.~C.~Lee}
\affiliation{Physics Department, Gangneung-Wonju National University, Gangneung, Republic of Korea}

\author{C.~Leroy}
\affiliation{D\'{e}partement de physique, Universit\'{e} de Montr\'{e}al, Qu\'{e}bec,  Canada}

\author{G.~Levi} 
\affiliation{INFN, Section of Bologna \& Department of Physics \& Astronomy, University of Bologna, Italy}

\author{A.~Lionti}
\affiliation{Section de Physique, Universit\'{e} de Gen\`{e}ve, Geneva, Switzerland}

\author{J.~Mamuzic}
\affiliation{IFIC, Universitat de Val\`{e}ncia - CSIC, Valencia, Spain}

\author{A.~Margiotta}
\affiliation{INFN, Section of Bologna \& Department of Physics \& Astronomy, University of Bologna, Italy}

\author{N.~Mauri}
\affiliation{INFN, Section of Bologna, Bologna, Italy}

\author{N.~E.~Mavromatos}
\affiliation{Theoretical Particle Physics \& Cosmology Group, Physics Dept., King's College London, UK}

\author{P.~Mermod}
\email[Corresponding author:\\]{philippe.mermod@cern.ch}
\affiliation{Section de Physique, Facult\'e des Sciences, Universit\'{e} de Gen\`{e}ve, Geneva, Switzerland}

\author{V.~A.~Mitsou}
\affiliation{IFIC, Universitat de Val\`{e}ncia - CSIC, Valencia, Spain}

\author{R.~Orava}
\affiliation{Physics Department, University of Helsinki, Helsinki, Finland}

\author{I.~Ostrovskiy}
\affiliation{Department of Physics and Astronomy, University of Alabama, Tuscaloosa, Alabama, USA}

\author{B.~Parker}
\affiliation{Institute for Research in Schools, Canterbury, UK}


\author{L.~Patrizii}
\affiliation{INFN, Section of Bologna, Bologna, Italy}

\author{G.~E.~P\u{a}v\u{a}la\c{s}}
\affiliation{Institute of Space Science, Bucharest - M\u{a}gurele, Romania}

\author{J.~L.~Pinfold}
\affiliation{Physics Department, University of Alberta, Edmonton, Alberta, Canada}

\author{V.~Popa}
\affiliation{Institute of Space Science, Bucharest - M\u{a}gurele, Romania}

\author{M.~Pozzato}
\affiliation{INFN, Section of Bologna, Bologna, Italy}

\author{S.~Pospisil}
\affiliation{IEAP, Czech Technical University in Prague, Czech~Republic}

\author{A.~Rajantie}
\affiliation{Department of Physics, Imperial College London, UK}

\author{R.~Ruiz~de~Austri}
\affiliation{IFIC, Universitat de Val\`{e}ncia - CSIC, Valencia, Spain}

\author{Z.~Sahnoun}
\affiliation{INFN, Section of Bologna, Bologna, Italy}
\affiliation{Centre for Astronomy, Astrophysics and Geophysics, Algiers, Algeria}

\author{M.~Sakellariadou}
\affiliation{Theoretical Particle Physics \& Cosmology Group, Physics Dept., King's College London, UK}

\author{A.~Santra}
\affiliation{IFIC, Universitat de Val\`{e}ncia - CSIC, Valencia, Spain}

\author{S.~Sarkar}
\affiliation{Theoretical Particle Physics \& Cosmology Group, Physics Dept., King's College London, UK}

\author{G.~Semenoff}
\affiliation{Department of Physics, University of British Columbia, Vancouver, British Columbia, Canada}

\author{A.~Shaa}
\affiliation{Formerly at Department of Physics and Applied Physics, Nanyang Technological University, Singapore  -- Associate member}

\author{G.~Sirri}
\affiliation{INFN, Section of Bologna, Bologna, Italy}

\author{K.~Sliwa}
\affiliation{Department of Physics and Astronomy, Tufts University, Medford, Massachusetts, USA}

\author{R.~Soluk}
\affiliation{Physics Department, University of Alberta, Edmonton, Alberta, Canada}

\author{M.~Spurio}
\affiliation{INFN, Section of Bologna \& Department of Physics \& Astronomy, University of Bologna, Italy}

\author{Y.~N.~Srivastava}
\affiliation{Physics Department, Northeastern University, Boston, Massachusetts, USA}


\author{M.~Suk}
\affiliation{IEAP, Czech Technical University in Prague, Czech~Republic}

\author{J.~Swain}
\affiliation{Physics Department, Northeastern University, Boston, Massachusetts, USA}

\author{M.~Tenti}
\affiliation{INFN, CNAF, Bologna, Italy}

\author{V.~Togo}
\affiliation{INFN, Section of Bologna, Bologna, Italy}


\author{J.~A.~Tuszy\'{n}ski}
\affiliation{Physics Department, University of Alberta, Edmonton, Alberta, Canada}

\author{V.~Vento}
\affiliation{IFIC, Universitat de Val\`{e}ncia - CSIC, Valencia, Spain}

\author{O.~Vives}
\affiliation{IFIC, Universitat de Val\`{e}ncia - CSIC, Valencia, Spain}

\author{Z.~Vykydal}
\affiliation{IEAP, Czech Technical University in Prague, Czech~Republic}


\author{A.~Widom}
\affiliation{Physics Department, Northeastern University, Boston, Massachusetts, USA}

\author{G.~Willems}
\affiliation{Physics Department, University of Muenster, Muenster, Germany}

\author{J.~H.~Yoon}
\affiliation{Physics Department, Konkuk University, Seoul, Korea}

\author{I.~S.~Zgura}
\affiliation{Institute of Space Science, Bucharest - M\u{a}gurele, Romania}

\collaboration{THE MoEDAL COLLABORATION}
\noaffiliation

\date{\today}

\begin{abstract}
We update our previous search for trapped magnetic monopoles in LHC Run 2 using nearly six times more integrated luminosity and including additional models for the interpretation of the data. The MoEDAL forward trapping detector, comprising 222~kg of aluminium samples, was exposed to 2.11~fb$^{-1}$ of 13 TeV proton-proton collisions near the LHCb interaction point and analysed by searching for induced persistent currents after passage through a superconducting magnetometer. Magnetic charges equal to the Dirac charge or above are excluded in all samples. The results are interpreted in Drell-Yan production models for monopoles with spins 0, 1/2 and 1: in addition to standard point-like couplings, we also consider couplings with momentum-dependent form factors. The search provides the best current laboratory constraints for monopoles with magnetic charges ranging from two to five times the Dirac charge.
\end{abstract}


\maketitle

\section{Introduction}

The magnetic monopole is motivated by the symmetry between electricity and magnetism, by grand-unification theories~\cite{tHooft1974,Polyakov1974,Scott1980,Preskill1984}, and by the fundamental argument advanced by Dirac that its existence would explain why electric charge is quantised~\cite{Dirac1931}. Dirac's argument also predicts the minimum magnetic charge that a monopole should carry, the Dirac charge $g_D$, which is equivalent to 68.5 times the elementary electric charge. It follows that a relativistic ($\beta=\frac{v}{c}>0.5$) monopole would ionise matter at least a thousand times more than a relativistic proton or electron. The Dirac charge $g_{\rm D}$ is obtained considering the electron charge $e$ as the fundamental unit of free electric charge; it is worth noting though that using the down-quark charge $\frac{1}{3}e$ instead of $e$ results in a minimum magnetic charge of $3g_D$, although in this case one cannot apply the Dirac argument in its original form because quarks are confined~\cite{Preskill1984}. 

The monopole hypothesised by Dirac was assumed to be point-like and structureless, and as such its underlying microscopic theory is completely unknown. Monopoles with masses that could be in a range accessible to colliders have been predicted to exist within extensions of the standard model~\cite{Cho1997,Cho2015,Ellis2016,Arunasalam2017,Ellis2017}. Other potentially low-mass monopoles within grand-unification theories or string-inspired models have also been predicted recently~\cite{Kephart2017,Mavromatos2017}. However, these exhibit detailed structure and as a consequence their production by particle collisions is expected to be suppressed~\cite{Drukier1982}, although enhanced production might be expected in environments with strong magnetic fields and high temperatures, such as those characterising heavy-ion collisions~\cite{Gould2017,Gould2017b}. Our search for monopole production in high-energy proton-proton ($pp$) collisions directly probes for free stable massive objects carrying a single or multiple Dirac charges, without assumptions about the monopole's structure. Monopole pair-production cross-sections are constrained with some model dependence because the detector acceptance depends on the monopole energy and angular distributions. To extract mass limits and compare results from different experiments, in absence of a better approach, the custom is to use cross sections computed from specific pair-production models such as Drell-Yan (DY) at leading order, with the caveat that the coupling of the monopole to the photon is so large that perturbative calculations are not expected to be reliable. For this reason it is preferable to interpret the search using as many different but theoretically well predicated models as possible.

Direct searches for monopoles were performed each time a new energy regime was made available in a laboratory, including the CERN Large Electron-Positron (LEP) collider, the HERA electron-proton collider at DESY, and the Tevatron proton-antiproton collider at Fermilab, where direct pair production of monopoles was excluded up to masses of the order of $400$~GeV (assuming DY cross sections) for monopole charges in the range $1g_{\rm D}-6g_{\rm D}$~\cite{OPAL2008,Kalbfleisch2004,H12005,CDF2006}. To cover the broadest possible ranges of masses, charges and cross sections, LHC-based direct searches for monopoles ought to use several complementary techniques, including general-purpose detectors, dedicated detectors, and trapping~\cite{DeRoeck2012a}. Searches were made in data samples of 7 and 8~TeV $pp$ collisions at the LHC with the ATLAS and MoEDAL experiments, probing the TeV scale for the first time~\cite{ATLAS2012a,ATLAS2015a,MoEDAL2016}. As of 2015, multi-TeV masses can be probed in 13~TeV $pp$ collisions. The first direct constraints in this energy regime were set by an analysis of the MoEDAL forward trapping detector exposed to $0.371$~fb$^{-1}$ of $pp$ collisions in 2015~\cite{MoEDAL2017}, providing the best sensitivity to date in the range $2g_{\rm D}-5g_{\rm D}$. 

This paper presents a new search with the MoEDAL forward monopole trapping detector~\cite{MoEDAL2017}, using the same trapping array with both 2015 and 2016 exposures at LHC point-8. The integrated luminosity for 13 TeV $pp$ collisions, as measured by the LHCb collaboration~\cite{LHCb2014}, corresponds to $2.11\pm 0.02$~fb$^{-1}$. The trapping volume consists of 672 square aluminium rods with dimension 19$\times$2.5$\times$2.5~cm$^3$ for a total mass of 222~kg in 14 stacked boxes which were placed 1.62~m from the IP8 LHC interaction point under the beam pipe on the side opposite to the LHCb detector. The setup and conditions of exposure are identical to those used in the previous search~\cite{MoEDAL2017}. 


\section{Magnetometer measurements}
\label{magnetometer}

\begin{figure*}[tb]
  \begin{center}
    \includegraphics[width=0.99\linewidth]{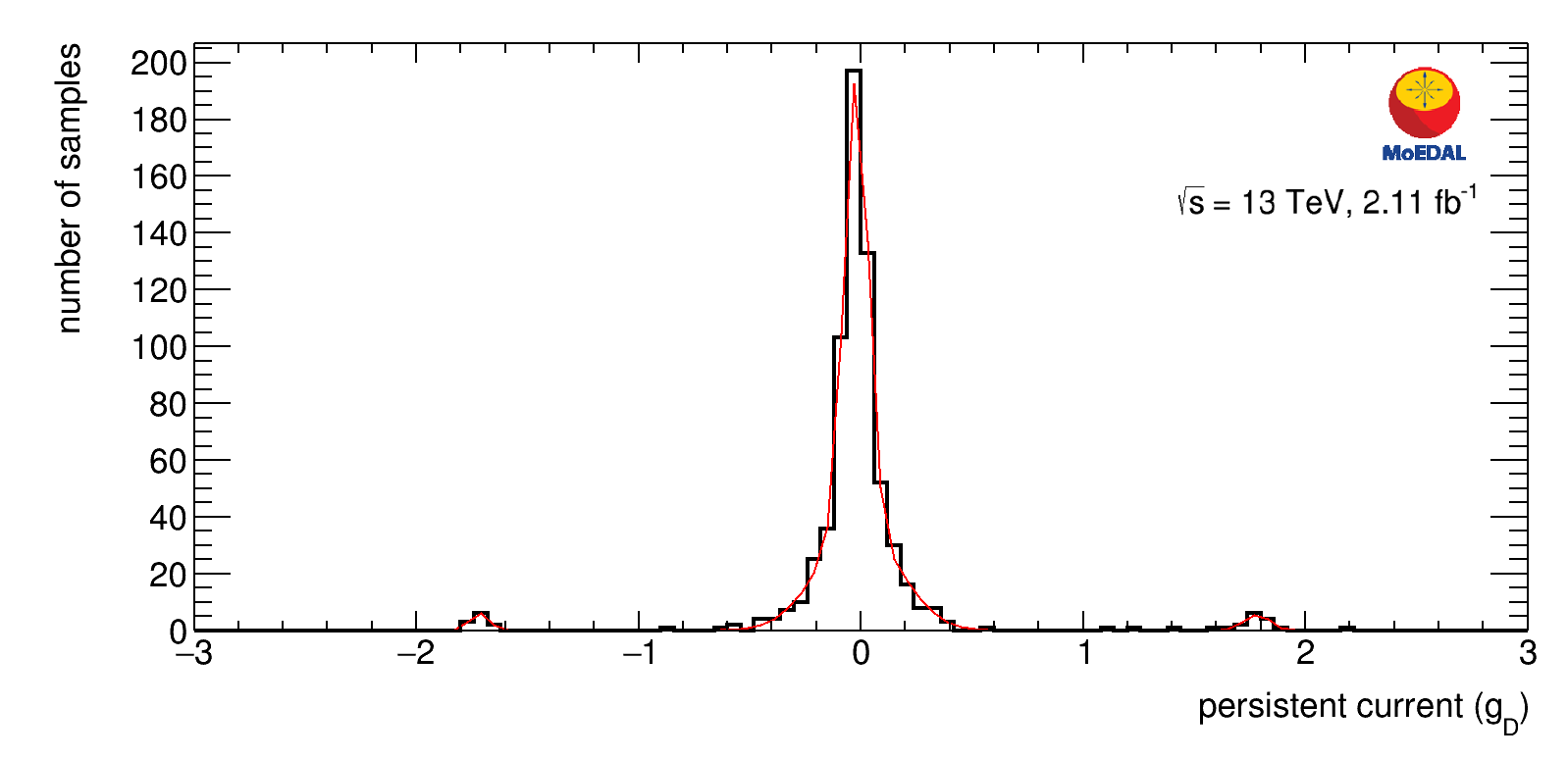}  
    \includegraphics[width=0.99\linewidth]{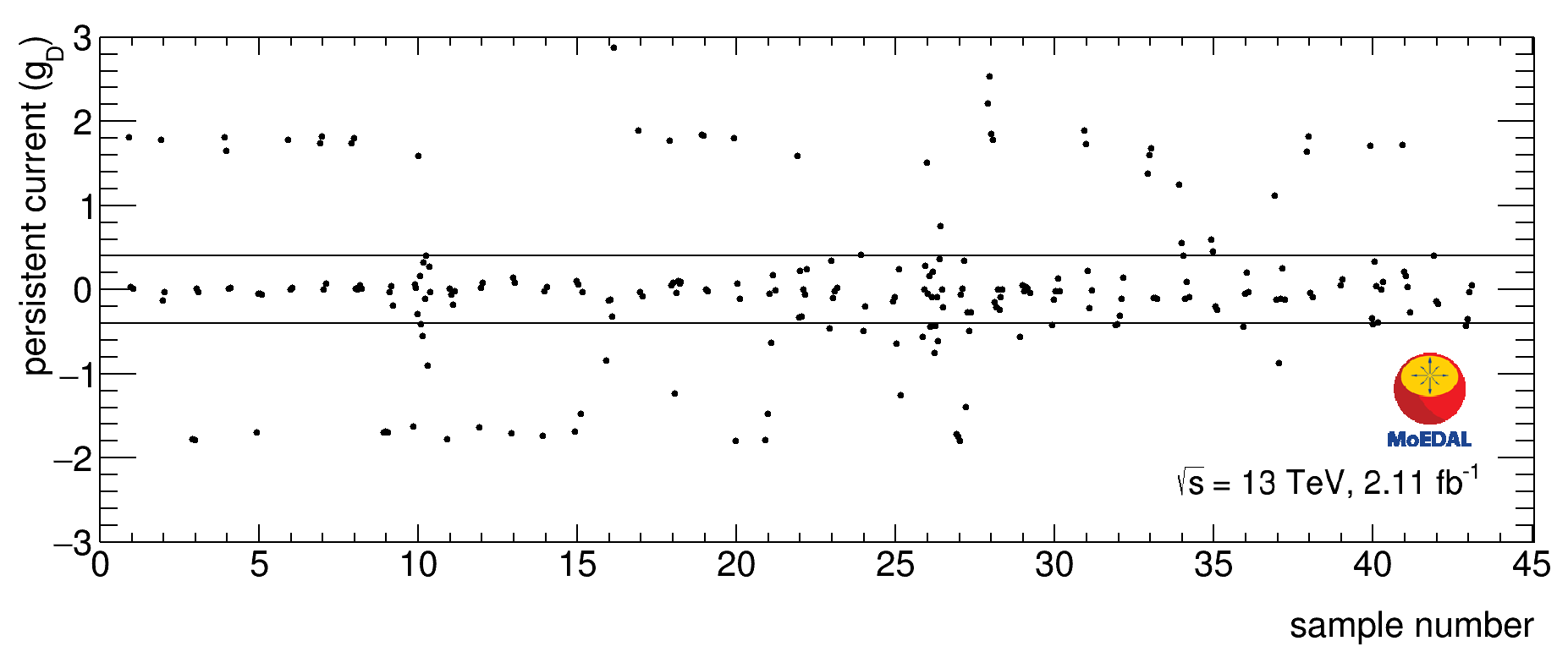}  
  \end{center}
  \caption{\label{fig:SQUID} Top: persistent current (in units of $g_{\rm D}$ after application of a calibration constant) after first passage through the magnetometer for all samples. The red curve shows a fit of the measured distribution using a sum of four Gaussian functions. Bottom: results of repeated measurements of candidate samples with absolute measured values in excess of $0.4g_{\rm D}$. }
\end{figure*}

\begin{figure}[tb]
  \begin{center}
    \includegraphics[width=1.\linewidth]{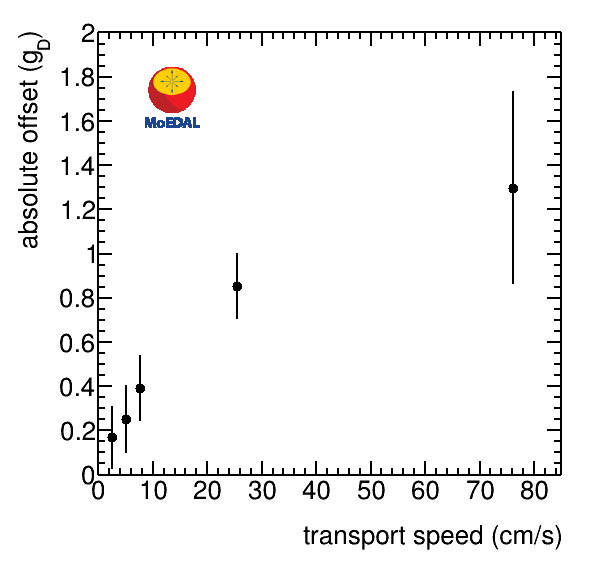}  
  \end{center}
  \caption{\label{fig:offset_vs_speed} Absolute value of the average persistent-current offset measured with magnetised calibration samples as a function of speed of transport through the magnetometer sensing region.}
\end{figure}

The 672 exposed aluminium samples of the MoEDAL forward trapping detector array were scanned in Spring 2017 during a two-week campaign with a DC SQUID long-core magnetometer (2G Enterprises Model 755) located at the laboratory for natural magnetism at ETH Zurich. Each sample was passed through the sensing coils at least once, with recordings of the magnetometer response in all three coordinates before, during, and after passage. The coil measuring the $z$ coordinate (along the shaft) is used for the monopole search because it circles the shaft in such a way that samples traverse it during transport; the $x$ and $y$ coordinates are used only to provide information about the strength and direction of magnetic dipole impurities contained in the samples. The persistent current is defined as the difference between the measured responses in the $z$ coordinate after and before passage of the sample, from which the contribution of the conveyor tray is subtracted. A calibration factor obtained from special calibration runs using two independent methods~\cite{DeRoeck2012b,MoEDAL2016,MoEDAL2017} is used to translate this value into the measured magnetic charge in the samples in units of Dirac charge. 

Persistent currents measured for all 672 samples for the first passage are shown in the top panel of Fig.~\ref{fig:SQUID}. Samples for which the absolute value of the measured magnetic charge exceeds a threshold of $0.4g_{\rm D}$ are set aside as candidates for further study. Measurements can then be repeated as many times as needed to minimise systematic errors and increase the sensitivity to the desired level. The $0.4g_{\rm D}$ threshold is chosen as a compromise between allowing sensitivity to magnetic charges down to $1g_{\rm D}$ and the time and effort required to scan a large number of samples multiple times. This gives 43 candidates, which were remeasured at least two more times each, as shown in the bottom panel of Fig.~\ref{fig:SQUID}. A sample for which a majority of measurements yields values below our threshold is considered to be a false positive, since a genuine monopole would consistently yield the same non-zero persistent current value.

During this measurement campaign, the identification of false positives was dominated by two effects. The first effect is a slight deterioration of the $z$ measurements due to random flux jumps occurring in the $x$ and $y$ measurements. Nearly all measurements in $z$ which result in a magnetic charge in the range $0.2g_{\rm D}<|g|<0.4g_{\rm D}$ are found to be correlated with a sudden jump in $x$ or $y$ (or both). The only effect of jumps in $x$ and $y$ was to add a new component to the resolution of the measurement in $z$, and recorded flux changes in the $z$ direction due to magnetic dipoles contained in the samples were observed to remain unaffected. The instabilities in the $x$ and $y$ sensors were found to be related to two phenomena: the build up of static charge on the sample tray while it moved along the track; and the capture of stray fields in the magnetometer. In Summer 2017, to mitigate these effects, an upgrade was performed which included the installation of an anti-static brush along the sample holder track, shielding of all cables between the SQUIDs and electronics, and grounding of the metallic magnetic shields. This was indeed observed to result in improved instrument performance and will be beneficial for future measurements. It also changed some of the conditions for test measurements performed after the upgrade to understand the second effect described below.

The second effect, which was already present in the previous runs~\cite{MoEDAL2017}, is an offset (generally taking a value around $\pm 1.8g_{\rm D}$) occasionally happening with samples containing magnetic dipole impurities. The mechanism causing the offset can be described as follows: whenever the sample magnetisation results in a flux inside the SQUID loop which temporarily exceeds the fundamental flux quantum $\Phi_0=\frac{h}{2e}$\footnote{This corresponds to $66g_{\rm D}$ for the $z$ direction taking into account the transfer from the pick-up coil and the specifications of the magnetometer.} within a given margin, the response may not return to the same level during the flux change in the other direction. This happens with magnetised samples regardless of exposure to LHC collisions, although the conditions for the offset to occur are not easy to control and reproduce. Valuable insights were provided by tests performed with calibration samples with magnetisation corresponding to the range $10^3g_{\rm D}-10^4g_{\rm D}$ when the sample is inside the sensing coils, for which frequent offsets were observed, confirming that the offset probability depends on the sample magnetisation. Moreover, an increased speed of transport through the sensing coils was observed to increase the offset value. Fig.~\ref{fig:offset_vs_speed} shows the average of the absolute value of the offset as a function of speed for magnetised calibration samples, confirming this trend and supporting the hypothesis that the offset is related to trapped fluxes inside the SQUID that occur when the slew rate (or rate of flux change) is increased~\cite{Clarke2006}. The offset value in the monopole search measurements was around $\pm 1.8g_{\rm D}$ for a transport speed of 5.08~cm/s. The same offset was observed for measurements performed with the calibration samples (Fig.~\ref{fig:offset_vs_speed}), although at higher speed, indicating that the offsets are mitigated by the magnetometer upgrade. In the present search, as can be seen in Fig.~\ref{fig:SQUID} (bottom), some candidates produced offsets more than once when remeasured. These particular candidates possess a higher magnetisation than average, corresponding to the range $10^2g_{\rm D}-10^3g_{\rm D}$ when the sample is inside the sensing coils. In each case, at least two measurements are consistent with zero magnetic charge. Moreover, the sign of the offset value in such candidates is reversed when turning the sample around such as to reverse its magnetisation in the $z$ direction. These features are consistent with the effect induced by the sample magnetisation described above and inconsistent with the presence of a monopole, confirming that all 43 candidates are false positives. 

Special care is given to the assessment of the probability for false negatives, i.e., the possibility that a monopole in a sample would remain unseen in the first pass due to a spurious fluctuation cancelling its response and resulting in a persistent current below the $0.4g_{\rm D}$ threshold used to identify candidates. This is studied using the distribution of persistent currents obtained in samples without monopoles, assuming that the magnetic field of the monopole itself (small compared to those of magnetic dipoles contained in the sample and tray) does not affect the mismeasurement probability. A template for this distribution is obtained from the search data themselves (top panel of Fig~\ref{fig:SQUID}) since we established from the multiple measurements that none of the candidates are genuine. A fit of this distribution ($\chi^2/\textrm{ndof}=0.74$) is obtained using a sum of four Gaussians: two Gaussians centred around zero to describe the shape of the main peak and the broadening of the resolution due to random flux jumps, respectively; and two Gaussians centred around $\pm 1.8g_{\rm D}$ to describe the occasional offsets. The probability to miss a monopole of charge $g$ is then estimated by integrating the fitted function in the interval $[-g-0.4g_{\rm D};-g+0.4g_{\rm D}]$ and dividing by the total number of samples (672). It is found to be less than 0.02\% for a magnetic charge $\pm 1g_{\rm D}$, less than 1.5\% for a magnetic charge $\pm 2g_{\rm D}$, and negligible for higher magnetic charges. These numbers could in principle be made even smaller by performing multiple measurements on all 629 non-candidate samples. However the level of detector efficiency obtained with the approach used is conservatively estimated to be 98\%. This is considered adequate for the search being performed and this efficiency is assumed for the final interpretation.

\section{Interpretation in pair-production models}

\begin{figure}[tb]
\begin{center}
  \includegraphics[width=0.6\linewidth]{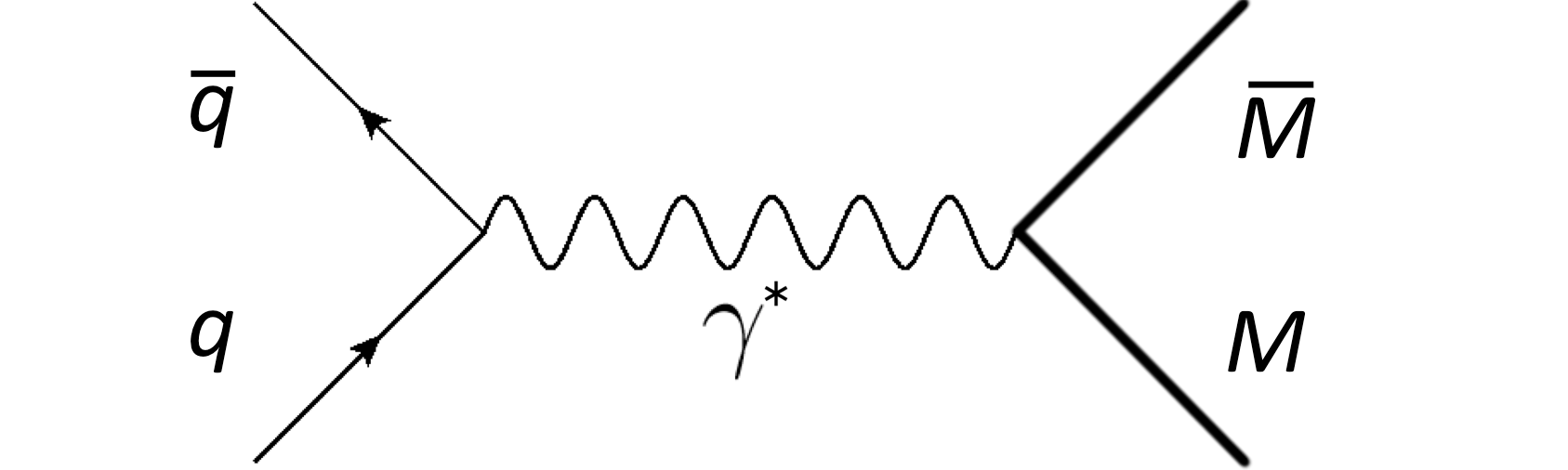}
  \caption{Feynman diagram for monopole pair production at leading order via the Drell-Yan process at the LHC. The non-perturbative nature of the process is ignored in the interpretation of the search.}
\label{fig:diagrams}
\end{center}
\end{figure}

The trapping detector acceptance, defined as the probability that a monopole of given mass, charge, energy and direction would end its trajectory inside the trapping volume, is determined from the knowledge of the material traversed by the monopole~\cite{LHCb2008,MoEDAL2016} and the ionisation energy loss of monopoles when they go through matter~\cite{Ahlen1978,Ahlen1980,Ahlen1982,Cecchini2016} implemented in a simulation based on G\textsc{eant}4~\cite{Geant42006}. For a given mass and charge, the pair-production model determines the kinematics and the overall trapping acceptance can be obtained. The uncertainty in the acceptance is dominated by uncertainties in the material description~\cite{MoEDAL2016,MoEDAL2017}. This contribution is estimated by performing simulations with material conservatively added and removed from the geometry model. 

\begin{figure*}[tb]
\begin{center}
  \includegraphics[width=0.495\linewidth]{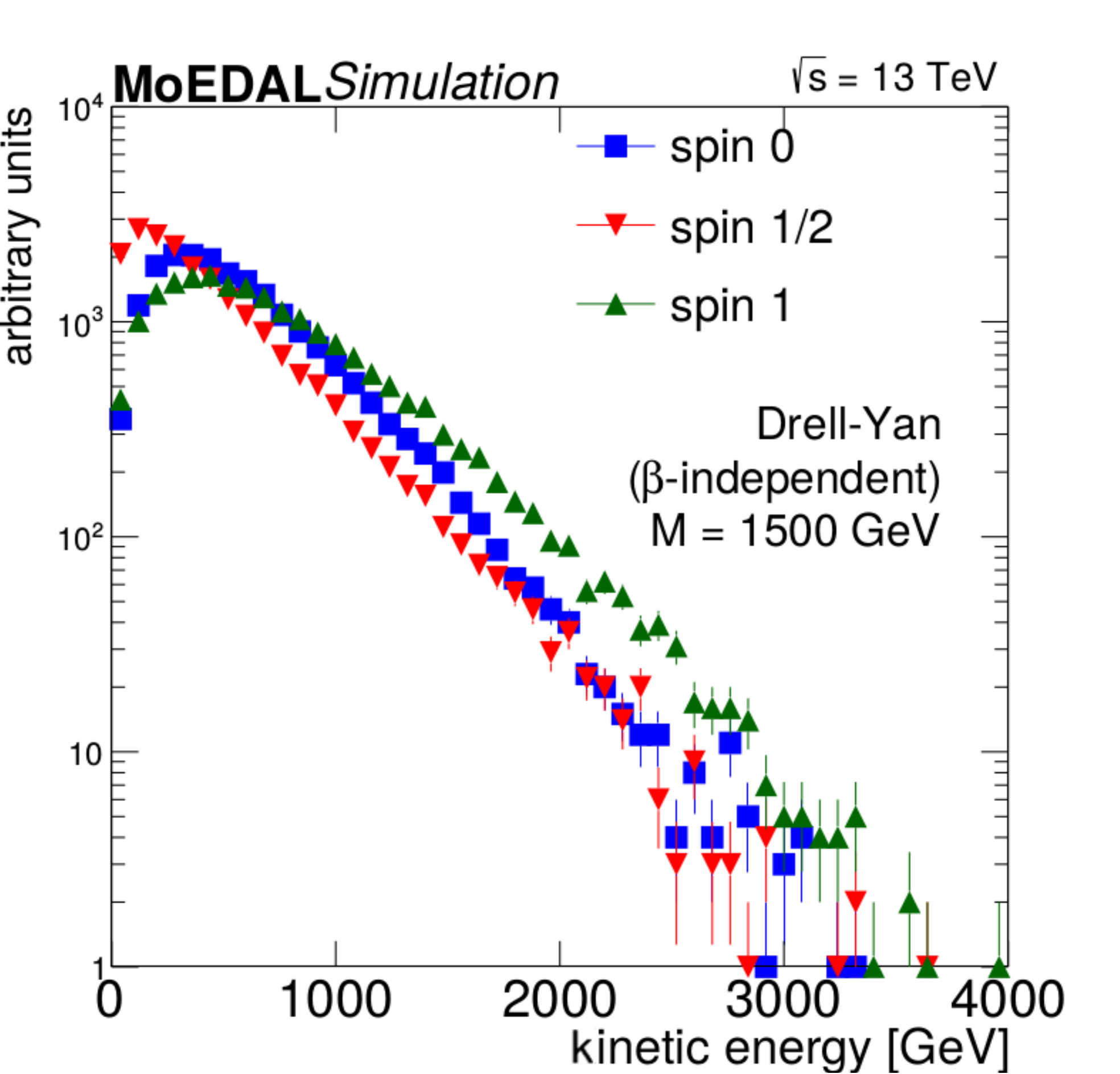}
  \includegraphics[width=0.495\linewidth]{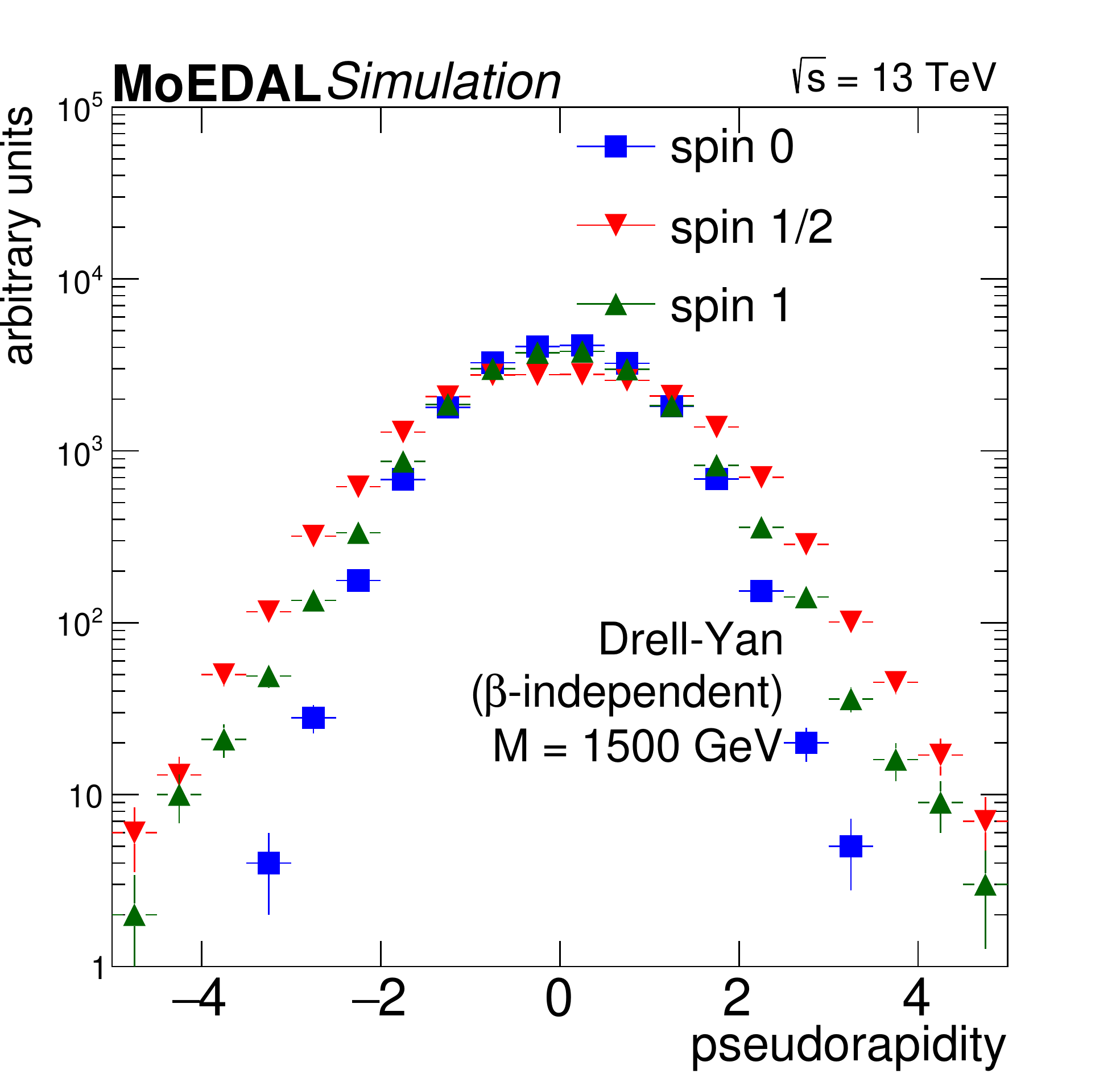}
  \includegraphics[width=0.495\linewidth]{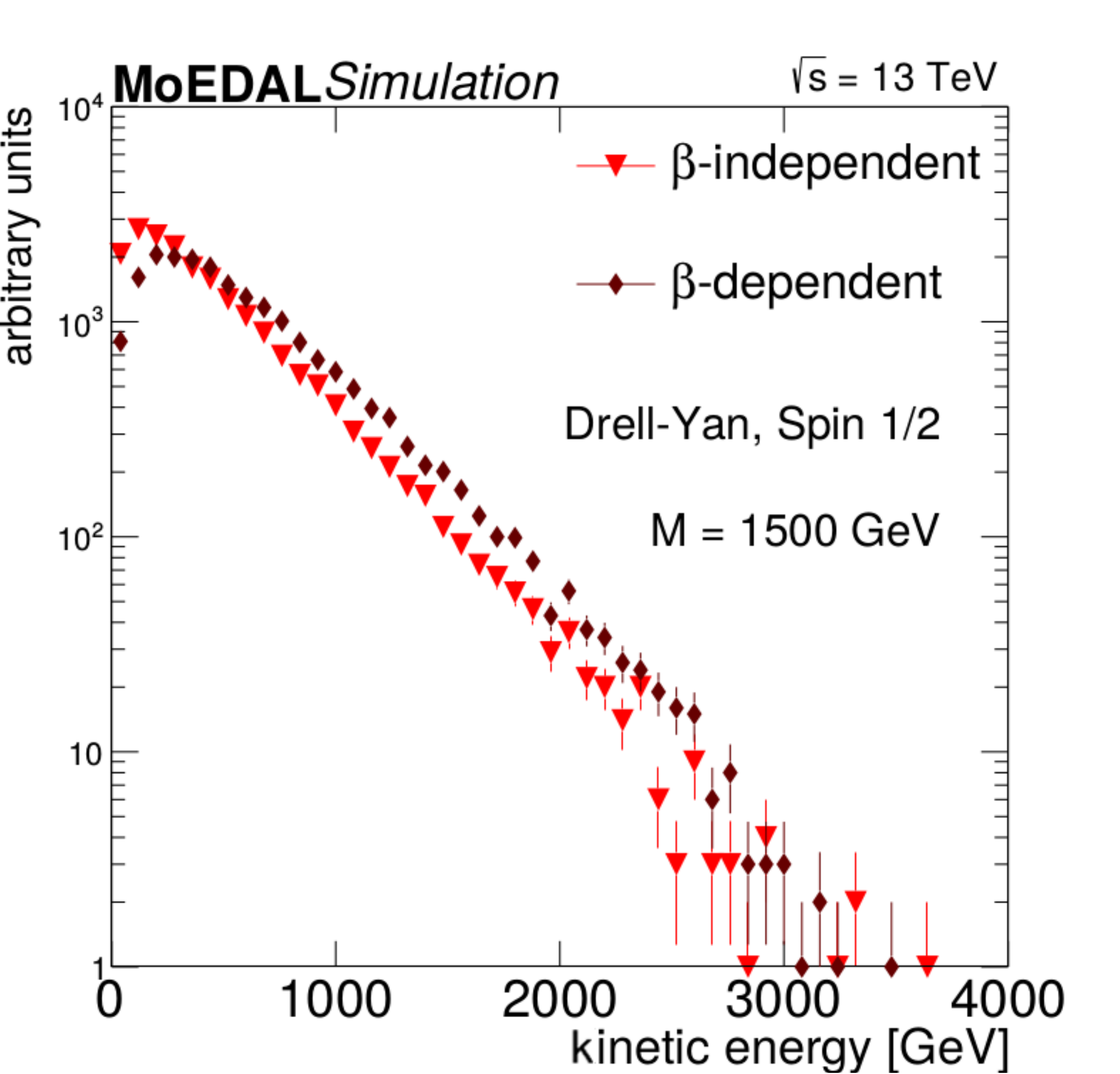}
  \includegraphics[width=0.495\linewidth]{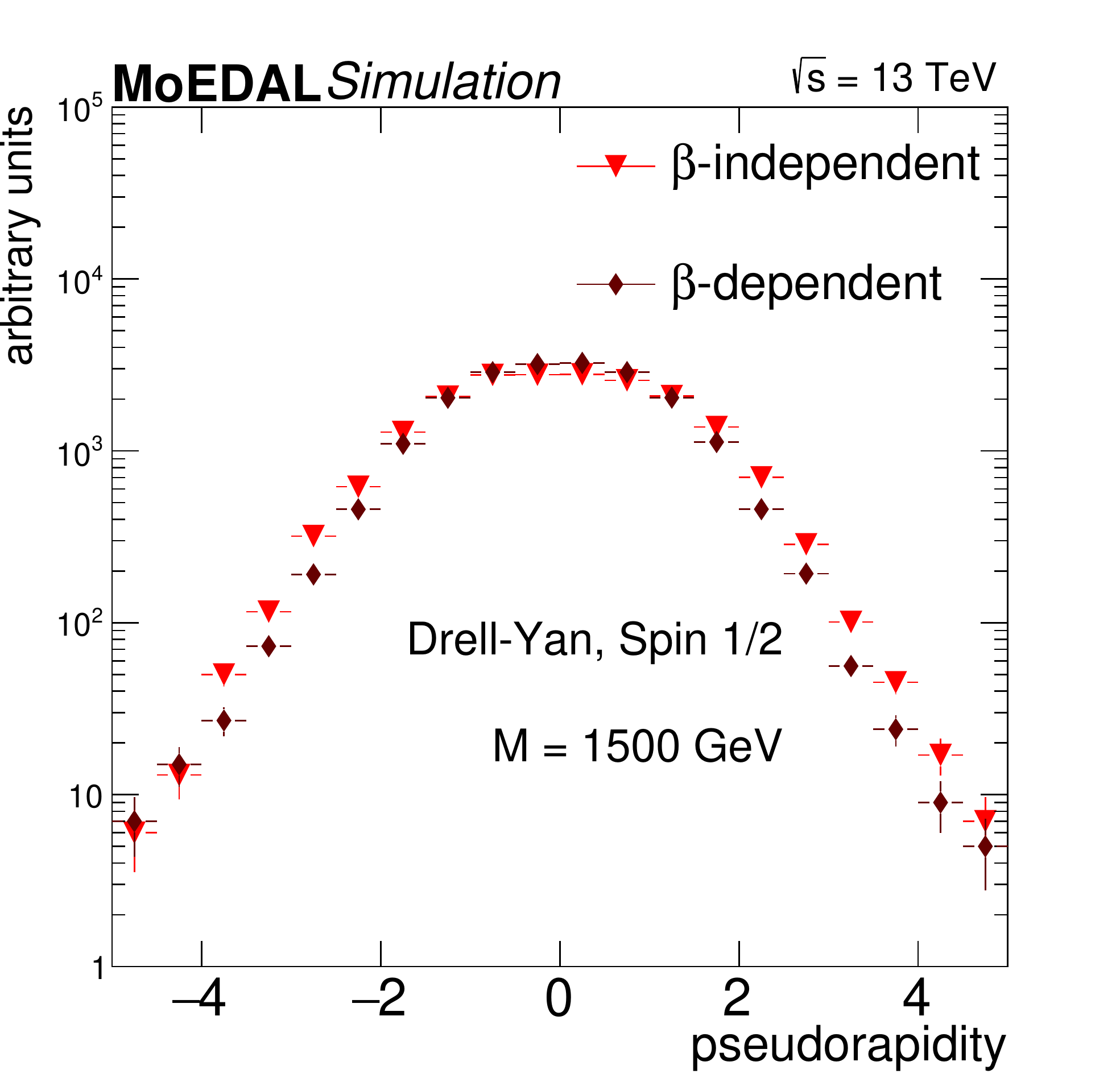}
  \caption{Distributions of kinetic energy (left) and pseudorapidity (right) for monopoles with mass 1500~GeV in models of Drell-Yan pair production generated by \MADGRAPH. The top plots show the standard $\beta$-independent coupling with different spin values (0, 1/2, 1) superimposed; and the bottom plots show spin-1/2 with two types of couplings ($\beta$-independent  and $\beta$-dependent) superimposed.}
\label{fig:DYBetaDist}
\end{center}
\end{figure*}

A Drell-Yan (DY) mechanism (Fig.~\ref{fig:diagrams}) is traditionally employed in searches to provide a simple model of monopole pair production~\cite{ATLAS2012a,ATLAS2015a,MoEDAL2016,MoEDAL2017}. In the interpretation of the present search, spin-1 monopoles are considered in addition to the spins 0 and 1/2 considered previously. The monopole magnetic moment is assumed to be zero and the coupling to the $Z$ boson is neglected. Models were generated in \MADGRAPH 5~\cite{Alwall2014} using only tree-level diagrams and the parton distribution function NNPDF2.3~\cite{Ball2013}. To extend the interpretation to a wider range of models, in addition to a point-like QED coupling, we also consider a modified photon-monopole coupling in which $g$ is substituted by $\beta g$ with $\beta = \frac{v}{c} = \sqrt{1 - \frac{4M^2}{s}}$ (where $M$ is the mass of the monopole and $\sqrt{s}$ is the invariant mass of the monopole-antimonopole pair), as was done at the Tevatron~\cite{Kalbfleisch2000,Kalbfleisch2004,CDF2006} and in the first ATLAS search~\cite{ATLAS2012a}. Such a modification, hereafter referred to as ``$\beta$-dependent coupling'', has been advocated in some studies~\cite{Schwinger1976,Gamberg2000,Milton2006,Kurochkin2006,Epele2012}, and illustrates the range of theoretical uncertainties in monopole dynamics. Using six different models for the interpretation of this search (three spin values and two kinds of coupling), with different angular and energy distributions as shown in Fig.~\ref{fig:DYBetaDist}, provides some measure of how the
choice of model affects the search acceptance. The reliability of all these models is limited no matter how, as current theories cannot handle the non-perturbative regime of strong magnetic couplings.


A comparison between the DY kinematic distributions when using different spin assumptions is shown in the top panels of Fig.~\ref{fig:DYBetaDist}. The observed differences are due to kinematic constraints imposed by angular momentum conservation. A comparison of $\beta$-independent and $\beta$-dependent photon-monopole coupling models is shown in the bottom panels. In the $\beta$-dependent case, a higher monopole energy is observed on average because the probability of generating a low-velocity monopole is suppressed by a factor $\beta<1$. 

The behaviour of the acceptance as a function of mass has two contributions: the mass dependence of the kinematic distributions, and the velocity dependence of the energy loss (lower at lower velocity for monopoles). For monopoles with $|g|=g_{\rm D}$, losses predominantly come from punching through the trapping volume, resulting in the acceptance being enhanced for a maximum of 3\% at low mass (high energy loss) and at high mass (low initial energy), with a minimum around 3000~GeV. The reverse is true for monopoles with $|g|>g_{\rm D}$ that predominantly stop in the upstream material and for which the acceptance is highest (up to 4\% for $|g|=2g_{\rm D}$, 2\% for $|g|=3g_{\rm D}$, and 1\% for $|g|=4g_{\rm D}$) for intermediate masses (around 1000~GeV). The acceptance remains below 0.1\% over the whole mass range for monopoles carrying a charge of $6g_{\rm D}$ or higher because they cannot be produced with sufficient energy to traverse the material upstream of the trapping volume. In this case the systematic uncertainties become too large and the interpretation ceases to be meaningful. The spin dependence is solely due to the different event kinematics. 

Simulations with uniform monopole energy distributions allow to identify, for various charge and mass combinations, ranges of kinetic energy and polar angle for which the acceptance is relatively uniform, called fiducial regions. The geometry of the setup used for this search is very similar to that of Ref.~\cite{MoEDAL2016} although with a slightly thicker trapping detector array. The fiducial regions given in this reference are expected to be identical except for the upper bounds in energy, which are generally not relevant because most monopoles in the collisions are produced at lower energies. They can thus conservatively be used to provide an interpretation which does not depend on the monopole production model. From the present search, a 95\% confidence level cross-section upper limit of 3.6~fb is set for monopoles produced in 13~TeV $pp$ collisions in the kinematic ranges of the fiducial regions which correspond to its mass and charge. 

\begin{figure*}[tb]
  \begin{center}
    \includegraphics[width=0.495\linewidth]{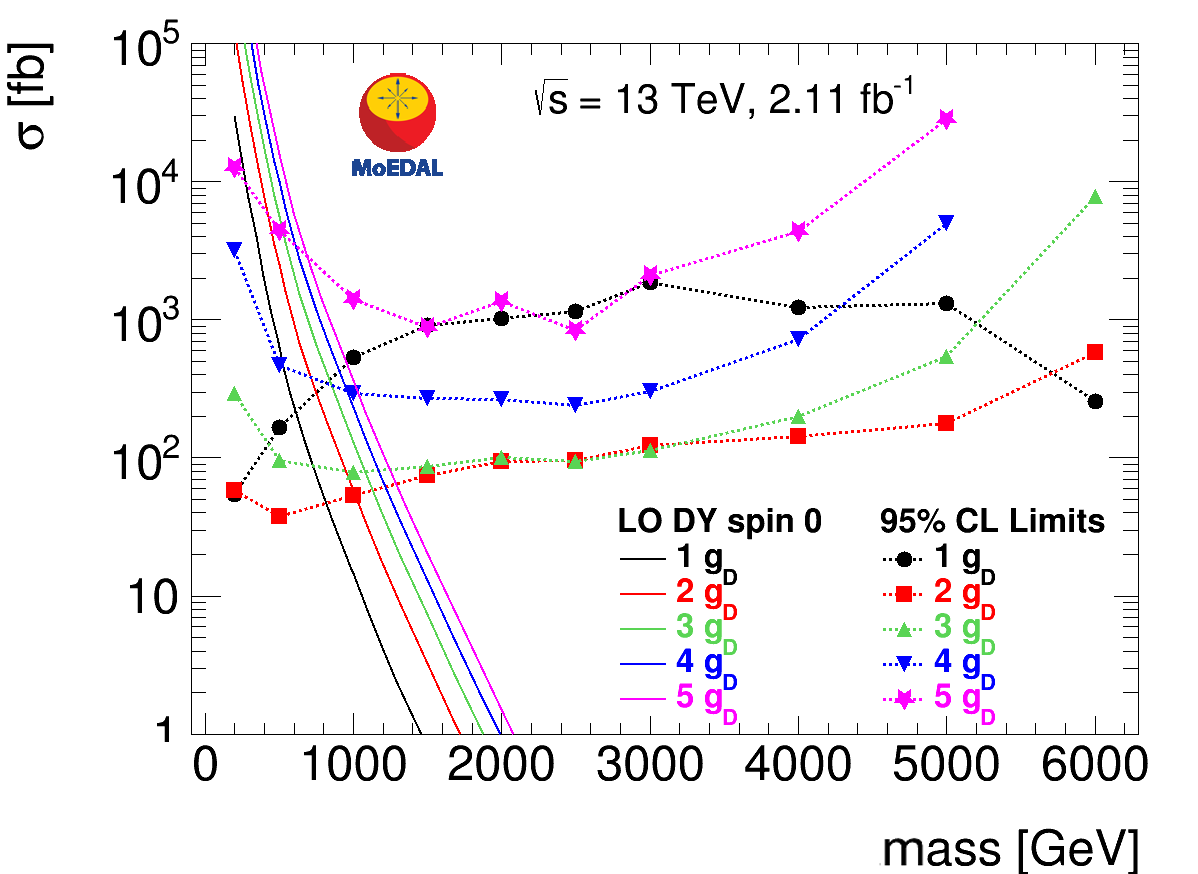}  
    \includegraphics[width=0.495\linewidth]{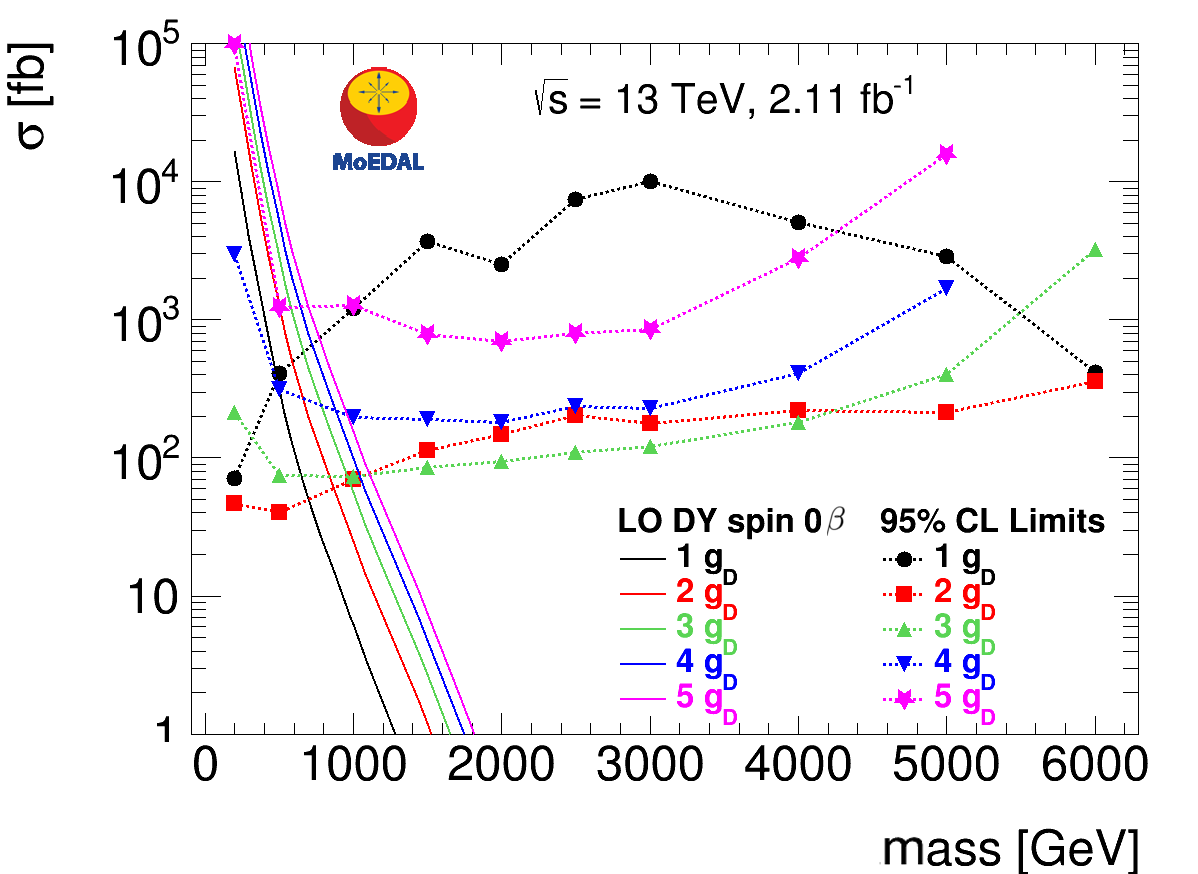}  
    \includegraphics[width=0.495\linewidth]{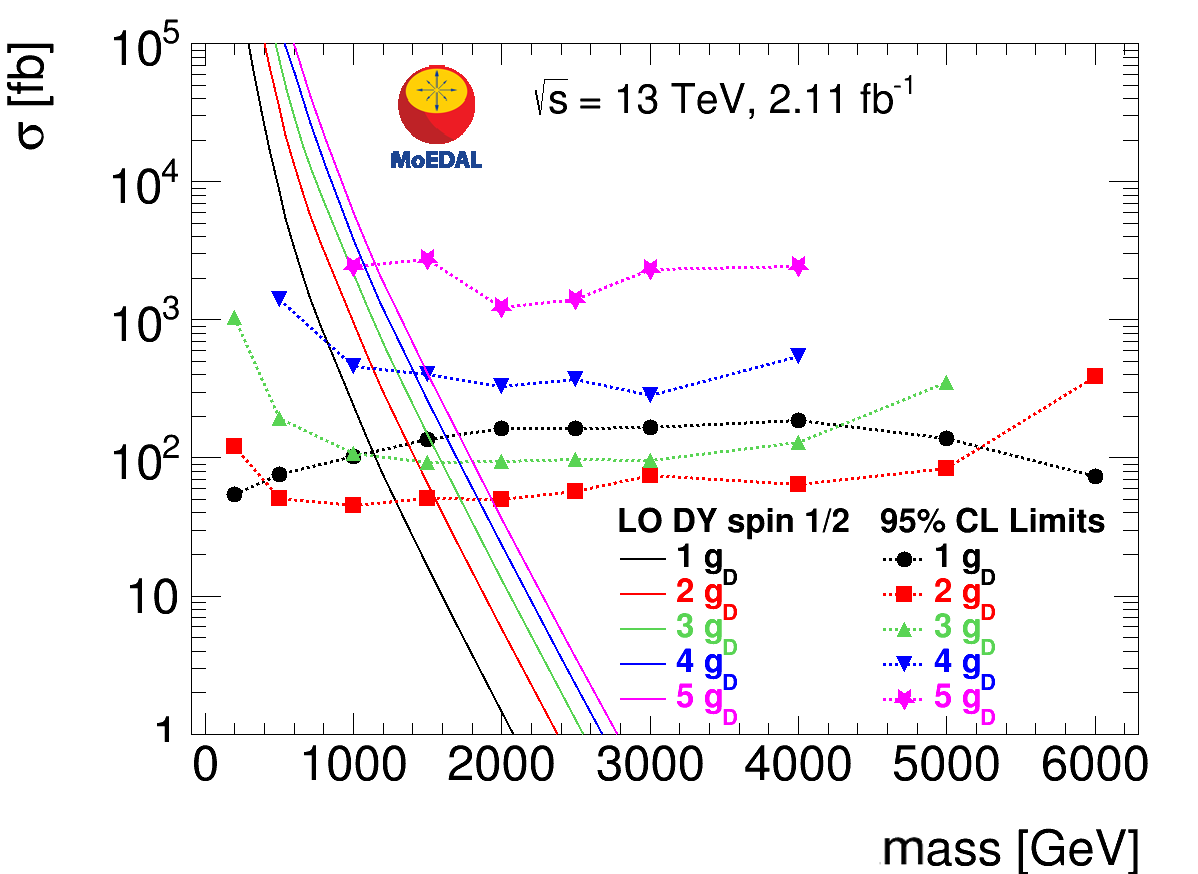}  
    \includegraphics[width=0.495\linewidth]{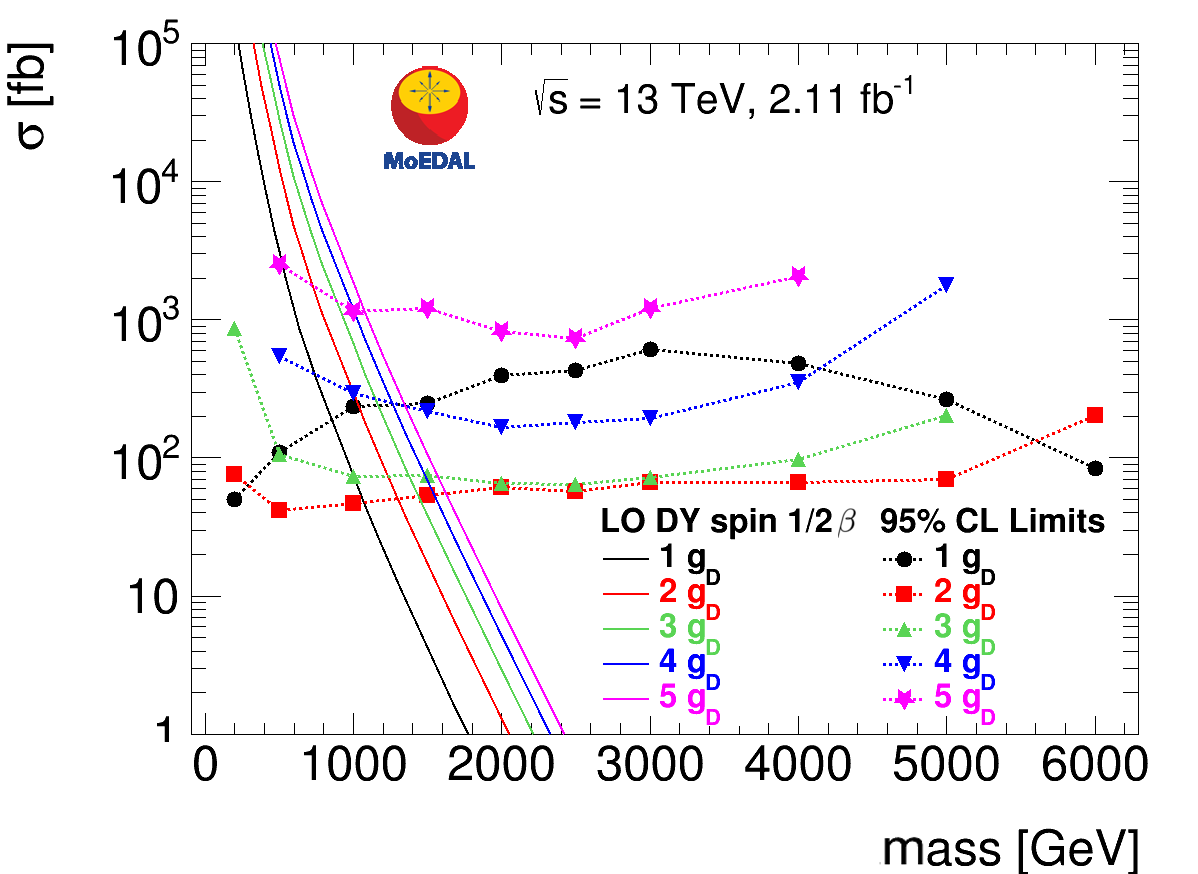}  
    \includegraphics[width=0.495\linewidth]{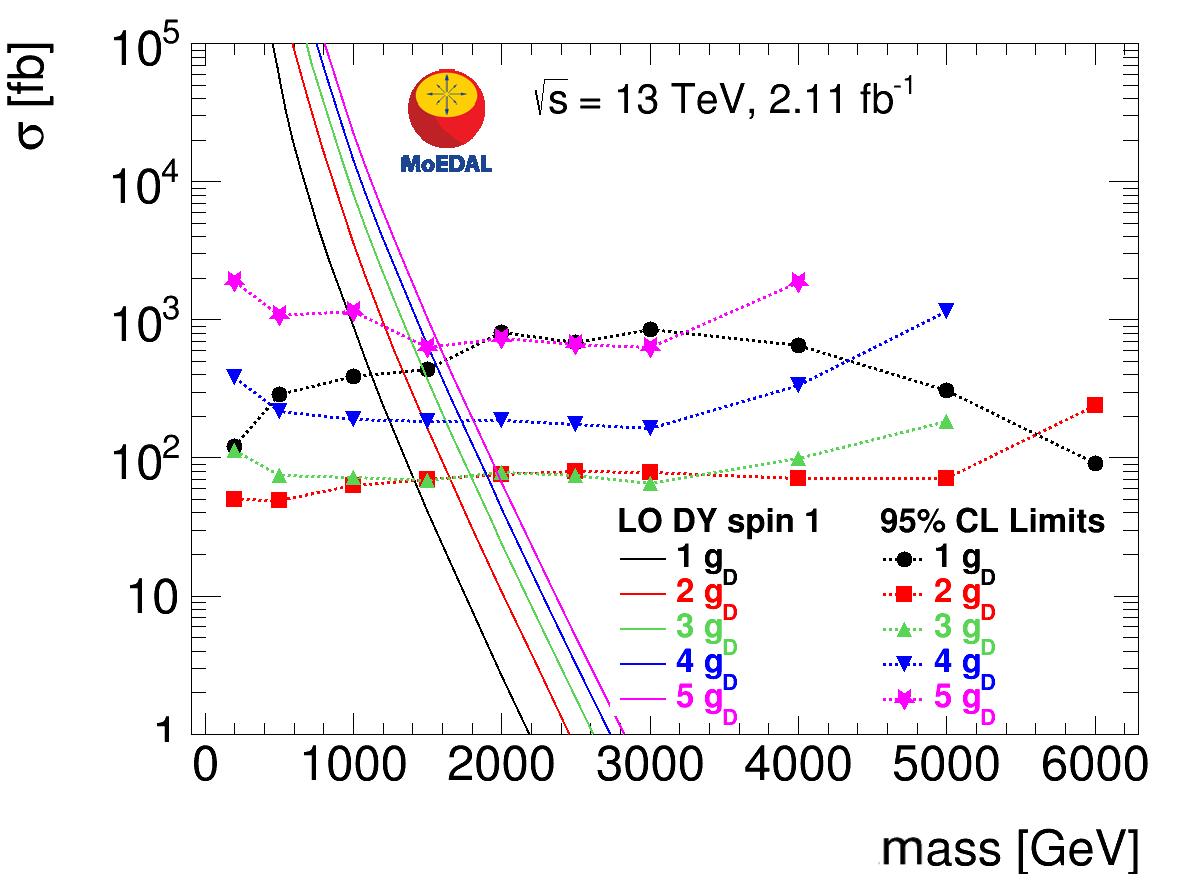}  
    \includegraphics[width=0.495\linewidth]{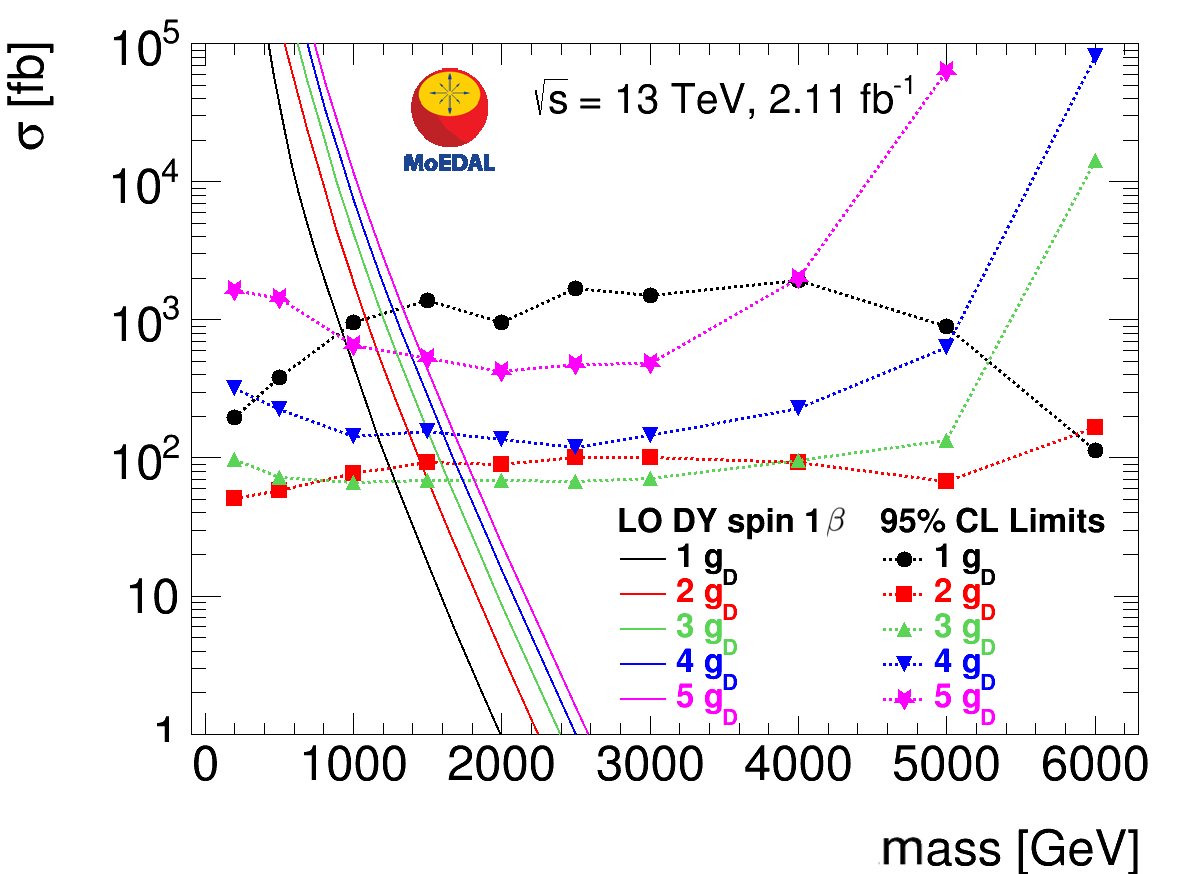}  
    \end{center}
  \caption{\label{fig:limits} Cross-section upper limits at 95\% confidence level for the DY monopole pair production model with $\beta$-independent (left) and $\beta$-dependent (right) couplings in 13 TeV $pp$ collisions as a function of mass for spin-0 (top), spin-1/2 (middle) and spin-1 (bottom) monopoles. The colours correspond to different monopole charges. Acceptance loss is dominated by monopoles punching through the trapping volume for $|g|=g_{\rm D}$ while it is dominated by stopping in upstream material for higher charges, explaining the shape difference. The solid lines are cross-section calculations at leading order (LO).}
\end{figure*}

Cross-section upper limits for DY monopole production with the two coupling hypotheses ($\beta$-independent, $\beta$-dependent) and three spin hypotheses (0, 1/2, 1) are shown in Fig.~\ref{fig:limits}. They are extracted from the knowledge of the acceptance estimates and their uncertainties; the integrated luminosity $2.11\pm 0.02$~fb$^{-1}$ corresponding to 2015 and 2016 exposure to 13~TeV $pp$ collisions; the expectation of strong binding to aluminium nuclei~\cite{Milton2006} of monopoles with velocity $\beta \le 10^{-3}$; and the non-observation of magnetic charge inside the trapping detector samples, with a 98\% efficiency (see Section~\ref{magnetometer}). 

Cross sections computed at leading order are shown as solid lines in Fig.~\ref{fig:limits}. Using these cross sections and the limits set by the search, indicative mass limits are extracted and reported in Table~\ref{tab:masslimits} for magnetic charges up to $5g_{\rm D}$. No mass limit is given for the spin-1/2 $5g_{\rm D}$ monopole with standard point-like coupling, because in this case the low acceptance at small mass does not allow MoEDAL to exclude the full range down to the mass limit set at the Tevatron of around 400~GeV for DY models~\cite{Kalbfleisch2004}.

\begin{table}[tb]
\begin{center}
\begin{tabular}{|l|c|c|c|c|c|}
\hline
Mass limits [GeV] & $1g_{\rm D}$ & $2g_{\rm D}$ & $3g_{\rm D}$  & $4g_{\rm D}$  & $5g_{\rm D}$ \\
\hline
MoEDAL 13 TeV &&&&& \\
(2016 exposure) &&&&& \\
 DY spin-0                           & 600      & 1000    & 1080  & 950    & 690   \\
 DY spin-\textonehalf                & 1110     & 1540    & 1600  & 1400   & --    \\
 DY spin-1                           & 1110     & 1640    & 1790  & 1710   & 1570  \\
 DY spin-0 $\beta$-dep.              & 490      & 880     & 960   & 890    & 690   \\
 DY spin-\textonehalf ~$\beta$-dep.  & 850      & 1300    & 1380  & 1250   & 1070  \\
 DY spin-1 $\beta$-dep.              & 930      & 1450    & 1620  & 1600   & 1460  \\
\hline
MoEDAL 13 TeV &&&&& \\
(2015 exposure) &&&&& \\
 DY spin-0               & 460      & 760      & 800   & 650   & --  \\
 DY spin-\textonehalf    & 890      & 1250     & 1260  & 1100  & --  \\
\hline
MoEDAL 8 TeV &&&&&  \\
 DY spin-0               & 420      & 600      & 560   & -- & -- \\
 DY spin-\textonehalf    & 700      & 920      & 840   & --  & -- \\
\hline
ATLAS 8 TeV &&&&&  \\
 DY spin-0               & 1050     & -- & -- & -- & -- \\
 DY spin-\textonehalf    & 1340     & -- & -- & -- & -- \\
\hline
\end{tabular}
\caption{95\% confidence level mass limits in models of spin-0, spin-1/2 and spin-1 monopole pair production in LHC $pp$ collisions. The present results (after $2016$ exposure) are interpreted for Drell-Yan
production with both $\beta$-independent and $\beta$-dependent couplings. These limits are based upon cross sections computed at leading order and are only indicative since the monopole coupling to the photon is too large to allow for perturbative calculations. Previous results obtained at the LHC are from Refs.~\cite{MoEDAL2016,MoEDAL2017} (MoEDAL in previous exposures) and Ref.~\cite{ATLAS2015a} (ATLAS).}
\label{tab:masslimits}
\end{center}
\end{table}

\section{Conclusions}

In summary, the aluminium elements of the MoEDAL trapping detector exposed to 13~TeV LHC collisions in 2015 and 2016 were scanned using a SQUID-based magnetometer to search for the presence of trapped magnetic charge. No genuine candidates were found. Consequently, monopole-pair direct production cross-section upper limits in the range $40-10^5$~fb were set for magnetic charges up to $5g_{\rm D}$ and masses up to 6~TeV. The possibility of spin-1 monopoles was considered for the first time in addition to spin-0 and spin-1/2, using a Drell-Yan pair-production model. Monopole mass limits in the range $490-1790$~GeV were obtained assuming cross sections at leading order -- the strongest to date at a collider experiment~\cite{PDG2016} for charges ranging from two to five times the Dirac charge. 

\section*{Acknowledgements}

We thank CERN for the very successful operation of the LHC, as well as the support staff from our institutions without whom MoEDAL could not be operated efficiently. We acknowledge the invaluable assistance of members of the LHCb Collaboration, in particular Guy Wilkinson, Rolf Lindner, Eric Thomas, and Gloria Corti. We thank Matt King, Rafal Staszewski and Tom Whyntie for their help with the software. We would like to acknowledge Wendy Taylor for her  invaluable input on the modelling of $\beta$-dependent couplings of magnetic monopoles. Computing support was provided by the GridPP Collaboration~\cite{GridPP2006,Britton2009}, in particular from the Queen Mary University of London and Liverpool grid sites. This work was supported by grant PP00P2\_150583 of the Swiss National Science Foundation; by the UK Science and Technology Facilities Council (STFC), via the research grants ST/L000326/1, ST/L00044X/1, ST/N00101X/1 and ST/P000258/1; by the Spanish Ministry of Economy and Competitiveness (MINECO), via the grants Grants No. FPA2014-53631-C2-1-P and FPA2015-65652-C4-1-R; by the Generalitat Valenciana via the special grant for MoEDAL CON.21.2017-09.02.03 and via the Projects PROMETEO-II/2014/066 and PROMETEO-II/2017/033; by the Spanish Ministry of Economy, Industry and Competitiveness (MINEICO), via the Grants No. FPA2014-53631-C2-1-P, FPA2014-54459-P, FPA2015-65652-C4-1-R and FPA2016-77177-C2-1-P; and by the Severo Ochoa Excellence Centre Project SEV-2014-0398; by the Physics Department of King's College London; by a Natural Science and Engineering Research Council of Canada via a project grant; by the V-P Research of the University of Alberta; by the Provost of the University of Alberta; by UEFISCDI (Romania); by the INFN (Italy); and by the Estonian Research Council via a Mobilitas Plus grant MOBTT5.

%

\bibliography{MMT2017}

\end{document}